\documentclass[12pt]{article}
\pdfoutput=1
\usepackage{epsf,amsfonts,amssymb,epsfig,amsmath,mathtools,graphics,slashed}
\usepackage{hyperref,hep,graphicx,subfig}
\usepackage{rotating}

\newcommand{\nn}{\notag \\}

\DeclareMathOperator{\Tr}{Tr}

\makeatletter

\@addtoreset{equation}{section}
\makeatother

\begin{document}

\begin{titlepage}

\vfill

\begin{flushright}
Imperial/TP/2016/JG/01\\
DCPT-16/27
\end{flushright}

\vfill

\begin{center}
   \baselineskip=16pt
   {\Large\bf Anisotropic plasmas from\\ axion and dilaton deformations}
  \vskip 1.5cm
  \vskip 1.5cm
Aristomenis Donos$^1$, Jerome P. Gauntlett$^2$ and Omar Sosa-Rodriguez$^1$\\
     \vskip .6cm
      \begin{small}
     \textit{$^1$Centre for Particle Theory\\ and Department of Mathematical Sciences\\Durham University\\Durham, DH1 3LE, U.K.}
        \end{small}\\\vskip .6cm
         \vskip .6cm
      \begin{small}
      \textit{$^2$Blackett Laboratory, 
  Imperial College\\ London, SW7 2AZ, U.K.}
        \end{small}\\

\end{center}

\vfill

\begin{center}
\textbf{Abstract}
\end{center}
\begin{quote}
We construct black hole solutions of type IIB supergravity that are holographically dual to anisotropic plasmas arising from deformations of an infinite class of four-dimensional CFTs. The CFTs are dual to $AdS_5\times X_5$, where $X_5$ is an Einstein manifold, and the deformations involve the type IIB axion and dilaton, with non-trivial periodic dependence
on one of the spatial directions of the CFT. 
At low temperatures the solutions approach smooth domain wall solutions with the same 
$AdS_5\times X_5$ solution appearing in the far IR.
For sufficiently large deformations an intermediate scaling regime appears which is
governed by a Lifshitz-like scaling solution.
We calculate the DC thermal conductivity and some components of the shear viscosity tensor.
\end{quote}

\vfill

\end{titlepage}

\setcounter{equation}{0}
\section{Introduction}
An interesting arena for studying holography is provided by
the $AdS_5\times X_5$ class of solutions of type IIB supergravity, where $X_5$
is a compact five-dimensional Einstein manifold. When $X_5=S^5$ the four-dimensional dual conformal theory (CFT) is given by $N=4$ supersymmetric Yang-Mills theory \cite{Maldacena:1997re}. Furthermore, when $X_5=SE_5$, where $SE_5$ is Sasaki-Einstein, the four-dimensional CFT has $N=1$ supersymmetry and, 
starting with the work of \cite{Klebanov:1998hh,Gauntlett:2004yd,Benvenuti:2004dy}, there 
are infinite classes of examples where both the geometry is explicitly known and the dual
field theory has also been identified.

The vacua of these theories have a complex modulus, given by constant values of $\tau\equiv\chi+ie^{-\phi}$, where 
$\chi$ and $\phi$ are the type IIB axion and dilaton, respectively,
corresponding to the fact that $\tau$ is dual to a marginal operator in the dual field theory. For the case of $N=4$ supersymmetric Yang-Mills theory $\tau$ is simply identified with the theta angle, $\theta$, and coupling constant, $g_{YM}$, via $\tau\sim\frac{\theta}{2\pi}+i\frac{4\pi}{g^2_{YM}}$. The identification for theories with $N=1$ supersymmetry is a little less direct due to renormalisation scales and is discussed in \cite{Klebanov:1998hh}.

In this paper we will be interested in studying how the entire class of CFTs 
behave under deformations of $\tau$, with the deformation having a non-trivial dependence on the spatial coordinates of the dual field theory. Such deformations trigger interesting RG flows which can, moreover, lead to novel IR ground states. Such spatially dependent deformations of CFTs are also of potential interest in seeking 
applications of holography with condensed matter systems. The deformations break translation invariance and hence they provide
a mechanism to dissipate momentum in the field theory. This leads to finite 
DC conductivities and hence these deformations can be used to model various metallic and insulating behaviour (e.g. \cite{Hartnoll:2012rj,Horowitz:2012ky,Donos:2012js,Donos:2013eha,Andrade:2013gsa,Donos:2014uba,Gouteraux:2014hca}).

An interesting setup, first studied in \cite{Azeyanagi:2009pr}, involves a deformation in which
the axion is linear in one of the spatial directions. It was shown that 
a novel ground state solution appears in the far IR of the RG flow which exhibits a spatially anisotropic Lifshitz-like scaling.
These solutions were generalised to finite temperature in \cite{Mateos:2011ix,Mateos:2011tv}, thus obtaining a dual description of a strongly coupled homogeneous
and spatially anisotropic plasma. One motivation for studying such plasmas is that they may provide some insights into the quark-gluon plasma that is seen in heavy ion collisions.

For the particular case of $N=4$ SYM it has recently been shown \cite{Banks:2015aca} that the anisotropic plasma found in \cite{Mateos:2011ix,Mateos:2011tv}
undergoes a finite temperature phase transition, demonstrating that the Lifshitz-like
scaling ground state found in \cite{Azeyanagi:2009pr} is actually not realised in the far IR. A subsequent study  
\cite{Banks:2016fab} revealed additional phase transitions and it is not yet clear
what the true ground state is. However, it remains plausible that the Lifshitz ground state is the true ground state for linear axion deformations which are associated with infinite classes of $SE_5$ and other
$X_5$.

From a technical viewpoint the solutions involving linear axions are appealing because they can be obtained by numerically solving a system of ODEs. This is also a feature of another class of deformations by $\tau$ that were studied in \cite{Jain:2014vka} involving a linear dilaton. For this class it was shown that at $T=0$ the RG flow approaches an $AdS_4\times R$ solution\footnote{This ground state is  unstable for the case of $X_5=S^5$ \cite{Jain:2014vka} and so it would be interesting to examine the associated finite temperature phase transitions generalising
\cite{Banks:2015aca,Banks:2016fab}.} in the IR. 
Note that unlike the linear axion solutions, the dilaton becomes 
large in the linear dilaton solutions and hence string perturbation theory will break down for a non-compact spatial direction.
Both the linear axion and the linear dilaton solutions can be viewed as special examples of ``Q-lattices", introduced
in \cite{Donos:2013eha}, where one exploits a global symmetry of the gravitational theory in order to break translation invariance using a bulk matter sector while preserving translation invariance in the metric. For deformations involving $\tau$ the relevant bulk global symmetry is the $SL(2,R)$ symmetry of type IIB supergravity which acts on $\tau$ via fractional linear transformations.

Generic spatially dependent deformations of $\tau$ will lead to solutions, which we refer to as holographic $\tau$-lattices, that involve numerically solving PDEs. While it is certainly interesting to construct such solutions and explore their properties, it is natural to first ask if the linear axion and dilaton deformations exhaust the Q-lattice constructions. In fact they do not. The bulk global $SL(2,R)$ symmetry has three different conjugacy classes of orbits and,
as we will explain, the linear axion and the linear dilaton deformations
are associated with the parabolic and hyperbolic conjugacy classes, respectively. This leaves deformations associated with the elliptic conjugacy class that we study here.
We can easily obtain a simple ansatz for these deformations by changing coordinates in field space. Instead of the upper half plane, the new coordinates naturally parametrise the Poincar\'e disc and the
$\tau$-deformation of interest is obtained by taking a polar coordinate on the disc to
depend linearly, and hence periodically, on one of the spatial directions. 

The $\tau$-lattice solutions of \cite{Azeyanagi:2009pr,Mateos:2011ix,Mateos:2011tv,Jain:2014vka} and the new ones constructed here can all be found in a 
$D=5$ theory of gravity 
which
arises as a consistent KK truncation of type IIB supergravity on an arbitrary $X_5$. 
We will introduce this $D=5$ theory in section \ref{setup} where we will also briefly review the Lifshitz-like scaling solution found in \cite{Azeyanagi:2009pr}. The new holographic $\tau$-lattices will be presented in section \ref{htl}. The finite temperature solutions depend on two dimensionless parameters $T/k$ and $\lambda$, 
where $T$ is the temperature, while $k$ and $\lambda$ are the wave-number and
strength of the $\tau$-deformation, respectively. We show that at $T=0$ the new $\tau$-lattices all approach domain walls interpolating between $AdS_5\times X_5$ in the UV and the same $AdS_5\times X_5$ in the far IR, thus recovering full four-dimensional conformal invariance.
The underlying physical reason for this is that
the operator which we using to deform the CFT has vanishing spectral density at low energies for non-vanishing momentum. Similar domain walls have been seen in other settings involving deformations by marginal operators that break translation
symmetry \cite{Chesler:2013qla,Donos:2014gya} and, as in those examples, there is a renormalisation of relative length scales in moving from the UV to the IR.

A particularly interesting feature, for
large enough values of $\lambda$, is that the solutions have an intermediate scaling regime, 
governed by the Lifshitz-like scaling solution found in \cite{Azeyanagi:2009pr}. 
At finite temperature this intermediate scaling appears for a range of $T/k$ and we will show how it 
manifests itself in the temperature scaling of various physical quantities.
For
the $T=0$ domain wall solutions the intermediate scaling will appear for a range of the radial variable.
Since the Lifshitz-like scaling solution of \cite{Azeyanagi:2009pr} is singular
the $T=0$ domain wall RG flows can thus be viewed as a singularity resolving mechanism
somewhat similar to some other singularity resolving flows, both bottom-up \cite{Harrison:2012vy,Bhattacharya:2012zu,Kundu:2012jn,Bhattacharya:2014dea} and top-down \cite{Donos:2012yi}. 
An important difference is that here the RG flow is being driven by a deformation at non-vanishing momentum.

For the finite temperature plasmas we calculate some components of the shear viscosity tensor. The spin two components, $\eta_{||}$, with respect to the residual $SO(2)$ rotation symmetry, satisfy $4\pi \eta_{||}/s=1$, where $s$ is the entropy density,
as usual \cite{Policastro:2001yc,Kovtun:2004de}. However, the spin one
components $\eta_\perp$ behave differently.
Defining $F\equiv 4\pi \eta_\perp/s$, for $T/k>>1$ we find $F\to 1$ 
while for $T/k<<1$ we find that $F$ approaches a constant given by
the renormalised relative length scales in the IR. For intermediate values of $T/k$ we have $F<1$, as seen in other anisotropic examples e.g. \cite{Rebhan:2011vd,Mamo:2012sy,Critelli:2014kra,Jain:2014vka,Jain:2015txa} (see \cite{Erdmenger:2010xm} for an anisotropic example where $F>1$). 

We also calculate the DC thermal conductivity of the plasmas in the anisotropic direction. Using the results of
\cite{Donos:2014cya,Donos:2015gia,Banks:2015wha,Donos:2015bxe} this can be expressed in terms of black hole horizon data and we find that it
has the correct scaling associated with the different scaling regimes. 
For $T<<k$ the Boltzmann behaviour of the thermal resistivity, $\kappa^{-1}$, is Boltzmann suppressed, which is expected because of the absence of low-energy excitations supported at the lattice momentum in the infrared, $k_{IR}$
\cite{Hartnoll:2012rj}. By contrast in the intermediate scaling regime, the lattice deformation gives rise 
to the power law behaviour $\kappa\sim k^2(T/k)^{7/3}$.
It is interesting to contrast these features with examples where power law behaviour occurs at low energies due to
the non suppression of spectral density at finite momentum in the context of semi-local quantum critical points\cite{Hartnoll:2012rj,Hartnoll:2012wm}.

We conclude the paper with some comments in section \ref{disc} while appendix \ref{appa} discusses the connection between the $SL(2,R)$ conjugacy classes and Q-lattices.

\section{The setup}\label{setup}

Our starting point is $D=5$ Einstein gravity with a negative cosmological constant, coupled to a complex scalar, $\Phi$,
with action given by
\begin{align}\label{eq:lag}
S=\int dx^{5}\sqrt{-g}\,\left(R+12-2\,\frac{\nabla_{\mu}\Phi\,\nabla^{\mu}\Phi^{\ast}}{\left(1-\Phi\Phi^{\ast} \right)^{2}} \right)\,.
\end{align}
The corresponding equations of motion are given by
\begin{align}\label{eq:eom}
R_{\mu\nu}+4\,g_{\mu\nu}-2\,\frac{\nabla_{\mu}\Phi\,\nabla_{\nu}\Phi^{\ast}}{\left(1-\Phi\Phi^{\ast} \right)^{2}}&=0\notag\,,\\
\nabla_{\mu}\left[ \frac{\nabla^{\mu}\Phi}{\left(1-\Phi\Phi^{\ast} \right)^{2}}\right]-2\,\frac{\nabla_{\mu}\Phi\,\nabla^{\mu}\Phi^{\ast}}{\left(1-\Phi\Phi^{\ast} \right)^{3}}\Phi&=0\,.
\end{align}

The complex scalar $\Phi$ parametrises the unit Poincar\'e disc $SU(1,1)/U(1)\cong SL(2,R)$. It will be helpful
to introduce two other standard choices of coordinates for the scalar field manifold. For the first we write
\begin{align}\label{eq:uphalf}
\Phi=\frac{1+i\,\tau}{1-i\,\tau},\qquad \tau=\chi+i\,e^{-\phi}\,,
\end{align}
where $\phi$ is the dilaton, $\chi$ is the axion and $\tau$ parametrises
the upper half plane. 
The metric on the scalar field manifold then takes the form
\begin{align}
ds_{2}^{2}\equiv 2\frac{d\Phi\,d\Phi^{\ast}}{(1-\Phi \Phi^{\ast})^{2}}&=\frac{d\tau d\bar\tau}{2(Im\tau)^2}=
\frac{1}{2}\,(d\phi^{2}+e^{2\phi}\,d\chi^{2})\,.\label{eq:scalar_metric_polar1}
\end{align}

The second choice of coordinates
\begin{align}\label{eq:polar}
\Phi=\tanh\frac{\varphi}{2}\,e^{i\,\alpha}\,,
\end{align}
 resembles polar coordinates with $\varphi\geq0$ and $0\leq \alpha<2\pi$ parametrising a circle. The two coordinate systems are related through the non-linear field redefinition:
\begin{align}\label{eq:field_redef}
\chi&=\frac{\sinh\varphi \, \sin\alpha}{\cosh\varphi+\sinh\varphi\,\cos\alpha}\,,\notag \\
e^\phi&=\cosh\varphi+\sinh\varphi\,\cos\alpha\,.
\end{align}
In these coordinates the metric on the scalar manifold is given by
\begin{align}
ds_{2}^{2}=\frac{2}{(1-\Phi \Phi^{\ast})^{2}}\,d\Phi\,d\Phi^{\ast}&
=\frac{1}{2}\,(d\varphi^{2}+\sinh^{2}\varphi\,d\alpha^{2})\,.\label{eq:scalar_metric_polar}
\end{align}

The $D=5$ theory \eqref{eq:lag} is a consistent truncation
of type IIB supergravity on a general five dimensional Einstein manifold $X_{5}$. In this truncation the IIB dilaton and axion are
precisely $\chi$ and $\phi$, respectively, while the self-dual IIB five-form is proportional to the $D=5$ volume form plus the volume form of $X_5$. The consistency of the truncation means that 
any solution of \eqref{eq:eom} can be uplifted on an arbitrary $X_5$ to obtain an exact solution of type IIB supergravity. In particular, the unit radius $AdS_5$ solution, given by
\begin{align}\label{eq:ads5}
ds^{2}_{5}&=r^{2}\,(-dt^{2}+d\mathbf{x}^{2})+\frac{dr^{2}}{r^{2}}\,,\qquad
\Phi=0\,,
\end{align}
uplifts to the vacuum $AdS_5\times X_5$ solution type IIB supergravity
and is dual to a CFT in four spacetime dimensions. Thus, the solutions
of this paper are applicable to this infinite class of CFTs. For the special case when 
$X_5=S^5$ the CFT is $N=4$ supersymmetric Yang-Mills theory and when 
$X_5=SE_5$ the dual CFT has $N=1$ supersymmetry.

The complex scalar field $\tau$ is massless when expanded about the $AdS_5$ vacuum and is dual to an exactly marginal operator of the dual CFT. Indeed 
constant values
of $\tau$, which we write as $\tau^{(0)}\equiv \chi^{(0)}+ie^{-\phi^{(0)}},$ parametrise a complex moduli space of CFTs. 
For the special case when $X_{5}=S^{5}$, the operators
dual to $\chi$ and $e^{-\phi}$ are proportional to $\Tr F\wedge F$ and $\Tr F^{2}$  
in $N=4$ Yang-Mills, respectively, and furthermore, we also have a simple identification 
$\tau^{(0)}\sim \frac{\theta}{2\pi}+\frac{4\pi i}{g^{2}_{YM}}$. To be more precise using
the conventions in section 4 of \cite{Mateos:2011tv}, which in particular means that for the $AdS_5$ vacuum solution we take $\phi^{(0)}=0$ and identify, for
$N=4$ Yang-Mills theory,
\begin{align}\label{taumap}
\frac{1}{g_s}\tau^{(0)}=\frac{\theta}{2\pi}+\frac{4\pi i}{g^{2}_{YM}}\,,
\end{align}
where $g_s$ is the string coupling constant. For general
$X_5=SE_5$ there is a less direct map because the dual field theory is described  
more implicitly, but nevertheless the complex modulus can easily be identified as described
in \cite{Klebanov:1998hh} (see also \cite{Benvenuti:2005wi}).

It is interesting to ask what happens to the field theory when we make the deformation parameter $\tau^{(0)}$ depend on some or all of the spatial coordinates, ${\mathbf x}$, of
the dual field theory. At strong coupling this can be addressed by constructing holographic solutions in which the bulk $\tau$ field behaves as $\tau(\mathbf{x},r)\to \tau^{(0)}(\mathbf{x})$ as one approaches the $AdS_5$ boundary at $r\to \infty$.
For the case of $N=4$ Yang-Mills theory, this deformation corresponds to 
a spatially dependent $\theta$ and $g^2_{YM}$ via \eqref{taumap} (with $g_s$ still a constant).
The most general deformation parameter $\tau^{(0)}(\mathbf{x})$ would require the solution of a system of non-linear PDEs in the bulk. However, for a particular class of boundary deformations one can maintain enough homogeneity to reduce the problem to a system of ODEs involving only the radial coordinate $r$. Indeed since the scalar fields parametrise a group manifold
$SL(2,R)$ the bulk equations of motion have a global $SL(2,R)$ symmetry and
hence one can consider the Q-lattice constructions described in \cite{Donos:2013eha}.

As we explain in appendix \ref{appa} there are three different Q-lattice constructions,
associated with the three conjugacy classes of $SL(2,R)$. 
In each case the scalar fields trace out a one-dimensional orbit, parametrised by one of
the spatial coordinates, which we take to be $z$. For all three cases 
the associated metric is
anisotropic and given by
\begin{align}\label{eq:metric_ansatz}
ds^{2}_{5}=-U(r)\,dt^{2}+\frac{dr^{2}}{U(r)}+e^{2\,V_{1}(r)}\,dz^{2}+e^{2\,V_{2}(r)}\,(dx^{2}+dy^{2})\,,
\end{align}
with all metric components functions of $r$ only.
The hyperbolic conjugacy class is associated with a linear dilaton, namely $\chi=0$ and $\phi=kz$, and
these solutions were constructed in \cite{Jain:2014vka}. Solutions associated with
the  parabolic conjugacy class have the axion $\chi$ linear in $z$. These solutions were studied in \cite{Azeyanagi:2009pr,Mateos:2011ix,Mateos:2011tv} and some features will be briefly reviewed in the next subsection. The focus of this
paper is solutions associated with the elliptic conjugacy class in which the field $\alpha$, introduced in \eqref{eq:polar}, 
is linear in the $z$ coordinate and will be discussed in the following section.

\subsection{Brief review of linear axion solutions}
The solutions obtained in \cite{Azeyanagi:2009pr} have 
\begin{align}
\phi=\phi(r),\qquad \chi=a\,z\,,
\end{align}
where $a$ is a constant,
supplemented with the metric ansatz \eqref{eq:metric_ansatz}. It is simple to check that
this gives a consistent non-linear ansatz for the equations of motion, leading to a system of ODEs which can be solved numerically. At the $AdS_5$ boundary the dilaton
vanishes so that the only deformation parameter is given by $a$.

For the special case when $X_5=S^5$, 
corresponding to $N=4$ Yang-Mills theory, the linear axion corresponds to
a $\theta$ angle in \eqref{taumap} that is linear in the $z$ direction. Also, recalling that the $AdS_5\times X_5$ solutions arise from $D3$-branes sitting at the apex of the metric cone over $X_5$, the
configurations with $\chi=a\,z$ are associated with the addition of $D7$-branes that
are aligned along the $x,y$ directions as well as wrapping $X_5$ and are smeared
along the $z$ direction. In fact it was shown in \cite{Mateos:2011ix} that 
$a=(\lambda_{tH}/4\pi) (n_{D7}/N_c)$ where $n_{D7}$ is the uniform
density of D7-branes in the $z$-direction, $N_c$ is the number of $D3$-branes
and $\lambda_{tH}$ is the 't Hooft parameter of the dual field theory.

By analysing the system of ODEs it was shown in \cite{Azeyanagi:2009pr}
that for any deformation parameter $a$ there is an RG flow from $AdS_5$ in the UV to
a Lifshitz-like scaling solution in the far IR with
\begin{align}\label{eq:taka_ground}
U=\frac{12}{11}\,r^{2},\quad V_{1}=\frac{2}{3}\,\ln r,\quad V_{2}=\ln r,\quad e^{\phi}=r^{2/3}\,.
\end{align}
Notice that the metric of this IR solution admits the spatially anisotropic scaling symmetry $(t,x,y,z)\to (\nu t,\nu x,\nu y, \nu^{2/3}z)$ with $r\to \nu^{-1}r$. On the other hand
the dilaton changes under this scaling. This means that some but not all observables
will exhibit this scaling behaviour. The fact that we flow to a new fixed point in the IR demonstrates that
the linear axion deformation is marginally relevant\footnote{Note that this is also true of the linear
axion solutions constructed in \cite{Andrade:2013gsa} both at vanishing and non-vanishing charge density.}.

Finite temperature generalisations of these solutions were constructed and studied in some detail in \cite{Mateos:2011ix,Mateos:2011tv}. For the special case of uplifting these solutions on $X_5=S^5$, an important subtlety is that in type IIB supergravity there are perturbative modes around the IR geometry \eqref{eq:taka_ground} which are unstable \cite{Azeyanagi:2009pr} and hence, for this case, the true ground state cannot be described by the geometry \eqref{eq:taka_ground}.
The unstable modes lie in the ${\bf 20}^\prime$ representation of the global $SO(6)$ R-symmetry and the back reaction of some of them have been studied recently
using an enlarged consistent Kaluza-Klein truncation, 
valid just for the case when $X_5=S^5$, in \cite{Banks:2015aca,Banks:2016fab}. 
By constructing finite temperature solutions it was shown in \cite{Banks:2015aca}
that there is a phase transition at finite temperature leading to a low temperature ground state geometry which was again Lifshitz-like but with different scaling exponents \cite{Banks:2015aca}. However, subsequent work in \cite{Banks:2016fab} revealed additional instabilities and the current status is that the true ground state for this case is not yet clear.
On the other hand for generic $X_5$ manifolds the instability of 
\cite{Azeyanagi:2009pr} is not present and one can hope that \eqref{eq:taka_ground} will be the true ground state for an infinite sub-class of cases.

\section{A new holographic $\tau$-lattice}\label{htl}

We now turn our attention to a different non-linear ansatz associated with the parametrisation \eqref{eq:polar}. It is simple to check that the ansatz for the scalars
\begin{align}\label{eq:q_ansatz}
\varphi=\varphi(r),\quad  \alpha=k\,z\,,
\end{align}
combined  with the metric given by \eqref{eq:metric_ansatz} is consistent, leading to a system of
ODEs. In particular, the scalar equation of motion can be written
\begin{align}\label{sceom}
\left(e^{V_1+2V_2} U\varphi'   \right)'=\frac{1}{2}k^2e^{2V_2-V_1}\sinh2\varphi \,.
\end{align}

The boundary deformations are now given by $\lambda\equiv \varphi^{(0)}$, where $\varphi^{(0)}\equiv\varphi(r=\infty)$, and the period $k$. An important difference in this parametrisation of the complex scalar, compared with 
those in \cite{Jain:2014vka,Azeyanagi:2009pr},
 is that the boundary deformation is periodic in the coordinate $z$ with period $2\pi/k$. 
At fixed $r$, as 
we traverse a single period in the $z$ direction it is clear from \eqref{eq:polar} that we traverse a circle in the Poincar\'e disc centred at the origin and with radius $\tanh\frac{\varphi}{2}$. Equivalently,
in terms of $\chi,\phi$ 
from \eqref{eq:field_redef} this corresponds to a circle in the upper half plane centred on
the imaginary axis at $e^{-\phi}=(1+\tanh^2\frac{\varphi}{2})/(1-\tanh^2\frac{\varphi}{2})$
and with radius $2\tanh\frac{\varphi}{2}/(1-\tanh^2\frac{\varphi}{2})$.

Notice, in particular, for the case of $N=4$ Yang-Mills using \eqref{taumap}
we see that both couplings $\theta$ and $g^2_{YM}$ are modulated by the same period. We also notice that provided $\varphi$ is bounded in the bulk, which will turn out to be the case in the solutions we construct, then $e^{\phi}$ is also bounded and hence string perturbation theory is not breaking down for these solutions\footnote{Recall that we are using conventions where string perturbation theory is governed by $g_se^\phi$, where $g_s$ is a free constant.}.

Like the linear axion deformation we can again interpret these deformations as arising from $D3$-branes at the apex of the metric cone over $X_5$ with a distribution of $D7$-branes aligned along the $(x,y)$ directions, wrapping $X_5$ and smeared in the $z$-direction. A striking difference however is that the integral of $\chi^{(0)}$ along a period in the $z$-direction vanishes and so we have a distribution of both $D7$-branes
and anti $D7$-branes.

Before constructing numerical solutions, we first develop some intuition about what will happen with the solutions in the two limits $\lambda<<1$ and $\lambda>>1$.
For the small $\lambda$ limit it is enough to examine small and static fluctuations of the scalar based on the ansatz \eqref{eq:q_ansatz} around the $AdS_5$ vacuum 
solution \eqref{eq:ads5}. 
By linearising the scalar equation of motion \eqref{sceom} we deduce that
\begin{align}\label{eq:phi_pert}
\delta\phi(r)=\lambda\,\frac{k^{2}}{2\,r^{2}}\,K_{2}\left(\frac{k}{r}\right)\,,
\end{align}
where $K_2$ is a Bessel function,
which close to the $AdS$ boundary gives the desired falloff:
\begin{align}
\delta\varphi(r)=\lambda-\frac{\lambda\,k^{2}}{4\,r^{2}}+\frac{\lambda\,k^{4}}{64\,r^{4}}\,\left( 3-4\gamma-4\,\ln\left(\frac{k}{2\,r}\right)\right)+\cdots\,.
\end{align}
By expanding the perturbation \eqref{eq:phi_pert} close to the Poincar\'e horizon at $r=0$ we find
\begin{align}\label{eq:scalar_pert_nh}
\delta\varphi(r)=\lambda\,\sqrt{\frac{\pi}{8}}\,\left(\frac{k}{r}\right)^{3/2}\,e^{-k/r}+\cdots\,.
\end{align}
This perturbation will back react on the metric at order $\lambda^2$ and explicit expressions can be obtained in terms
of Meijer G-functions. The behaviour in \eqref{eq:scalar_pert_nh}
demonstrates that when $\lambda$ is small, the deformation of the boundary theory does not significantly affect the IR physics away from the $AdS_5$ vacuum \eqref{eq:ads5}. Indeed it is clear that $k$ sets a scale in the bulk with the geometry rapidly returning to the $AdS_5$ vacuum at $r<k$. At finite temperature, similar statements can be made for the corresponding horizon at temperatures $T<k$. 

We conclude that at $T=0$, at least for small $\lambda$, the deformation gives rise
to a domain wall solution interpolating between $AdS_5$ in the UV and the same $AdS_5$ in the IR.
This behaviour should be contrasted with what occurs for the linear axion deformations \cite{Azeyanagi:2009pr}
and the linear dilaton deformations \cite{Jain:2014vka},
both of which modify the lR.
As we will explain in more detail later there is a renormalisation of relative length scales
as one moves from the UV to IR, which for small 
$\lambda$ is of order $\lambda^2$. 
Note that similar domain walls have been shown to arise
in other contexts involving deformations with spatially dependent marginal operators \cite{Chesler:2013qla,Donos:2014gya}.

We now turn our attention to deformations corresponding to large $\lambda$. In this case, one has to construct the full geometry either in closed form or, as we do in the next section, numerically. However, we can still obtain some insight based on analytic arguments. When $\lambda$ is large, the scalar field $\varphi(r)$ close to the boundary of $AdS_{5}$ will also be large. In that region, the complex scalar target space metric in the $(\alpha,\varphi)$ coordinates given in \eqref{eq:scalar_metric_polar} can be approximated by
\begin{align}
ds^{2}_{2}\approx \frac{1}{2}\,\left(d\varphi^{2}+e^{2\varphi}\,d\left(\frac{\alpha}{2}\right)^{2} \right)\,,
\end{align}
which locally looks exactly like the metric \eqref{eq:scalar_metric_polar1} given in the $(\chi,\phi)$ coordinates. 
It is natural to expect, therefore, that for the zero temperature solutions there will be a large region where the metric scales according to \eqref{eq:taka_ground} while the scalar behaves as $\varphi\approx \frac{2}{3}\,\ln r$. As we then move deeper in the bulk geometry, the scalar becomes smaller and this approximation breaks down. 
It is then plausible that as soon as soon as we are in the region $r<k$, we move quickly back to the horizon of $AdS_{5}$ with the scalar behaving as in \eqref{eq:phi_pert}.

Similar comments apply at finite temperature. For high temperatures we expect the entropy density, $s$, will scale as $s\sim T^{3}$ corresponding to the scaling associated with the AdS-Schwarzschild black hole. This will also be the behaviour for $T<<k$ if the solution
approaches the $AdS_5$ to $AdS_5$ domain wall, which we know happens for small $\lambda$, and we will shortly see also happens for large $\lambda$. To be more precise, due to the length renormalisation we expect this behaviour for $T<<k_{IR}\equiv k/\bar L_1$ (with $\bar L_i$ defined below; see \eqref{lrenorm}).

In addition we expect 
an intermediate scaling region where $s$ would scale according to $s\sim T^{8/3}$
associated with the finite temperature version of the scaling solution \eqref{eq:taka_ground}. 
The lower bound of this region should satisfy $1<<T/k_{IR}$, to ensure that we are not dominated by the $T=0$ domain wall, 
while the upper bound will be 
fixed by ensuring that we are not dominated by the high $T$ AdS-Schwarzschild solution. A consideration of the
expansions of the functions in the UV that we give below \eqref{eq:uv_exp} suggests that we should have $T/k<<e^{\lambda}$.

\subsection{Numerical construction}\label{sec:num_construction}

We consider the ansatz for the metric given in \eqref{eq:metric_ansatz}
supplemented with the ansatz for the complex scalar $\Phi$ given by 
\eqref{eq:q_ansatz} and \eqref{eq:polar}. After substituting into the equations of motion
\eqref{eq:eom} we obtain a system of ODEs for 
four functions, $U,V_1,V_2$ and $\varphi$, of the radial coordinate $r$. 
The function $U$ satisfies a first order ODE, while the functions $V_{1}$, $V_{2}$ and $\varphi$ satisfy second order ones.
In order to find finite temperature solutions we use a standard double sided shooting technique which we now outline.

Close to the $AdS_{5}$ boundary, located at $r=\infty$, the expansion of the four functions has the form
\begin{align}\label{eq:uv_exp}
U(r) &= r^2 \left(1-\frac{k^2 \sinh ^2\lambda }{12 r^2}+ \frac{\mathcal{B}_1}{r^4}+\frac{\ln r}{72 r^4}k^4 \sinh ^2\lambda  (2 \cosh 2\lambda +1)+\cdots\right)\,, \nonumber\\
V_1 (r) &= \ln r+\frac{k^2 \sinh ^2(\lambda )}{12 r^2}+\frac{\mathcal{B}_2}{r^4}-\frac{\ln r}{72r^4}k^4 \sinh ^2\lambda  (2 \cosh 2\lambda +1)+\cdots\,, \notag\\
V_2 (r) &= \ln r -\frac{k^2 \sinh ^2\lambda }{24 r^2}+\frac{k^4 (\cosh 2\lambda -\cosh 4\lambda )-576\, \mathcal{B}_2}{1152 r^4}\nonumber\notag\\
\quad & \quad  \quad +\frac{\ln r}{144r^4}k^4 \sinh ^2\lambda  (2 \cosh 2\lambda +1)+\cdots\,, \notag\\
\varphi(r) &=  \lambda-\frac{k^2 \sinh 2\lambda }{8 r^2}+ \frac{\mathcal{B}_3}{r^4}-\frac{\ln r}{96 r^4}k^4 (\sinh 2\lambda -2 \sinh 4\lambda )+\cdots \,.
\end{align}
In particular, after fixing the scalar deformation parameter $\lambda$ as well as the length scales of the asymptotic metric, we are left with the three constants of integration $\mathcal{B}_{i}$. 
Notice that if we rescale the radial coordinate $r\to \nu r$ as well
as rescaling $(t,x,y,z)$ by $\nu^{-1}$ the ansatz will be preserved by the following scaling of the UV parameters:
\begin{align}\label{eq:uv_exp_scale}
k&\to \nu k,\qquad\lambda\to \lambda\,,\nn
{\mathcal{B}_1}&\to \nu^4{\mathcal{B}_1}-  \frac{(\nu k)^4}{72}\sinh ^2\lambda (2 \cosh 2\lambda +1)\ln \nu \,, \nonumber\\
{\mathcal{B}_2}&\to \nu^4 {\mathcal{B}_2}+\frac{(\nu k)^4}{72} \sinh ^2\lambda  (2 \cosh 2\lambda +1)\ln \nu\,, \notag\\
{\mathcal{B}_3}&\to \nu^4 {\mathcal{B}_3} + \frac{(\nu k)^4}{96} (\sinh 2\lambda -2 \sinh 4\lambda )\ln \nu\,.
\end{align}
The log terms are associated with anomalous scaling of physical quantities due to
the conformal anomaly which is non-vanishing (see \cite{Mateos:2011tv}
for a related discussion).

We also demand that the solutions have a regular black hole event horizon located at
some value $r=r_{+}$. This leads us to the following expansion near $r=r_+$:
\begin{align}\label{eq:nh_exp}
U &= h_{11}(r-r_{+})+ h_{12} (r-r_{+})^2+\dots\,,\nonumber\\
V_1 &= \mathcal{H}_2+\mathcal{H}_1 (r-r_{+})+h_{22} (r-r_{+})^2+\cdots\,,\nonumber\\
V_2 &= \mathcal{H}_3 + h_{31}(r-r_{+})+\cdots\,,\nonumber\\
\varphi &=  \mathcal{H}_4+h_{41}(r-r_{+})+\cdots\,. 
\end{align}
The expansion is specified by four free constants $\mathcal{H}_{a}$, with the remaining constants $h_{ij}$ fixed in terms of those. We find that the temperature and entropy density of the black holes are given by
\begin{align}
T &=\frac{r_+}{\pi}\,\frac{16 +k^2 (1-\cosh (2 \mathcal{H}_4))\,e^{-2 \mathcal{H}_2}}{16 (\mathcal{H}_1 r_+ +1)}\,,\notag\\
s &= 4 \pi  e^{\mathcal{H}_2+2 \mathcal{H}_3}\,,
\end{align}
respectively.

In total we have ten free constants: $\mathcal{B}_{i}$, $\lambda,k$, $\mathcal{H}_{a}$, and $r_+$, but one of these is redundant due to the scaling symmetry given in
\eqref{eq:uv_exp_scale}. By numerically solving the ODEs starting at both $r=r_+$ and
$r=\infty$ we match the four functions at some point in the middle $r=r_{m}$ along with continuity of the first derivatives $\varphi^{\prime}$, $V_{1}^{\prime}$ and $V_{2}^{\prime}$. This leads to seven conditions and thus the solution space is specified
by two dimensionless parameters which we take to be $\lambda$ and $T/k$.

\begin{figure}[t]
\centering
\includegraphics[height=4.8cm]{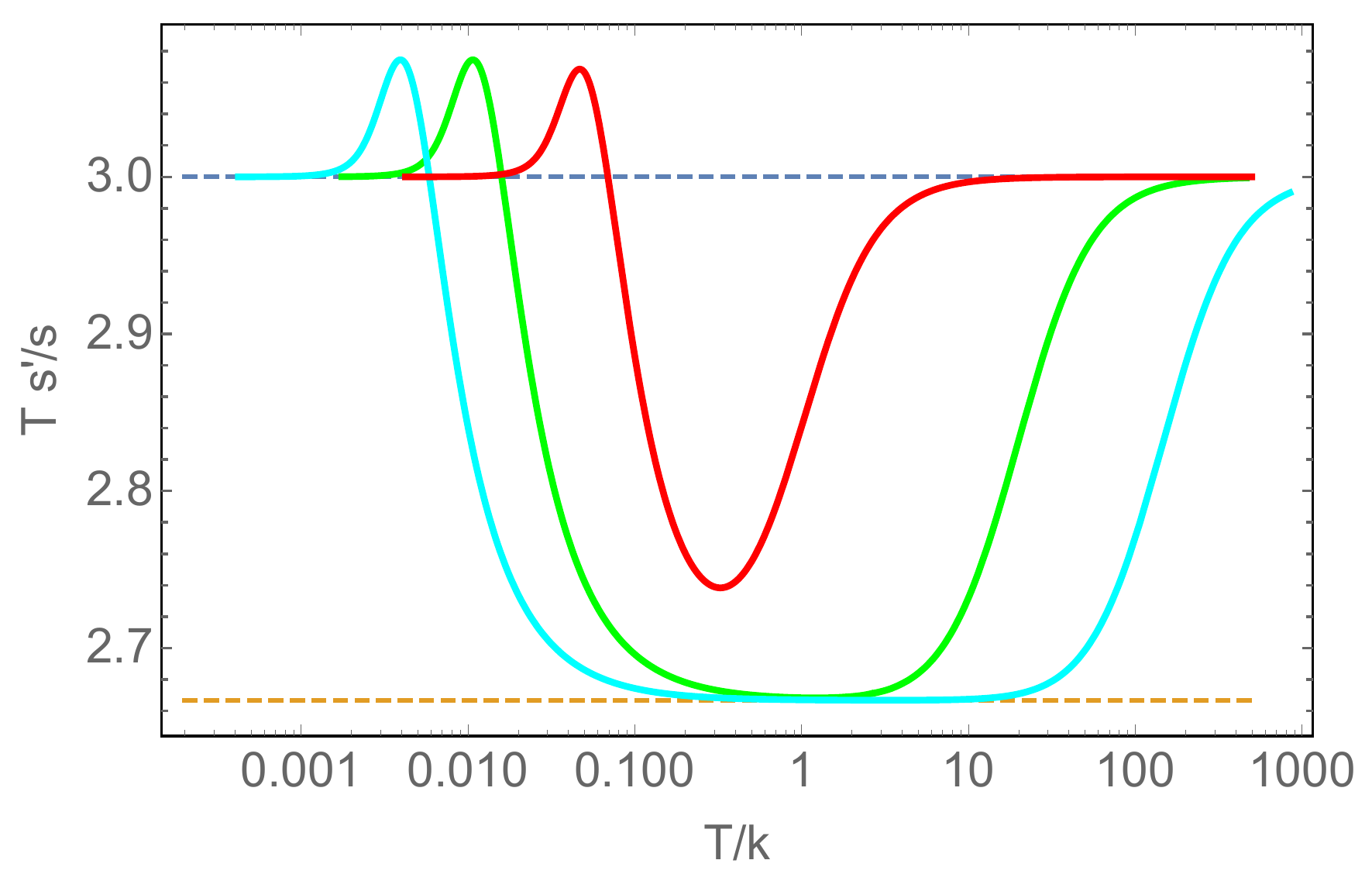}\quad
\includegraphics[height=4.8cm]{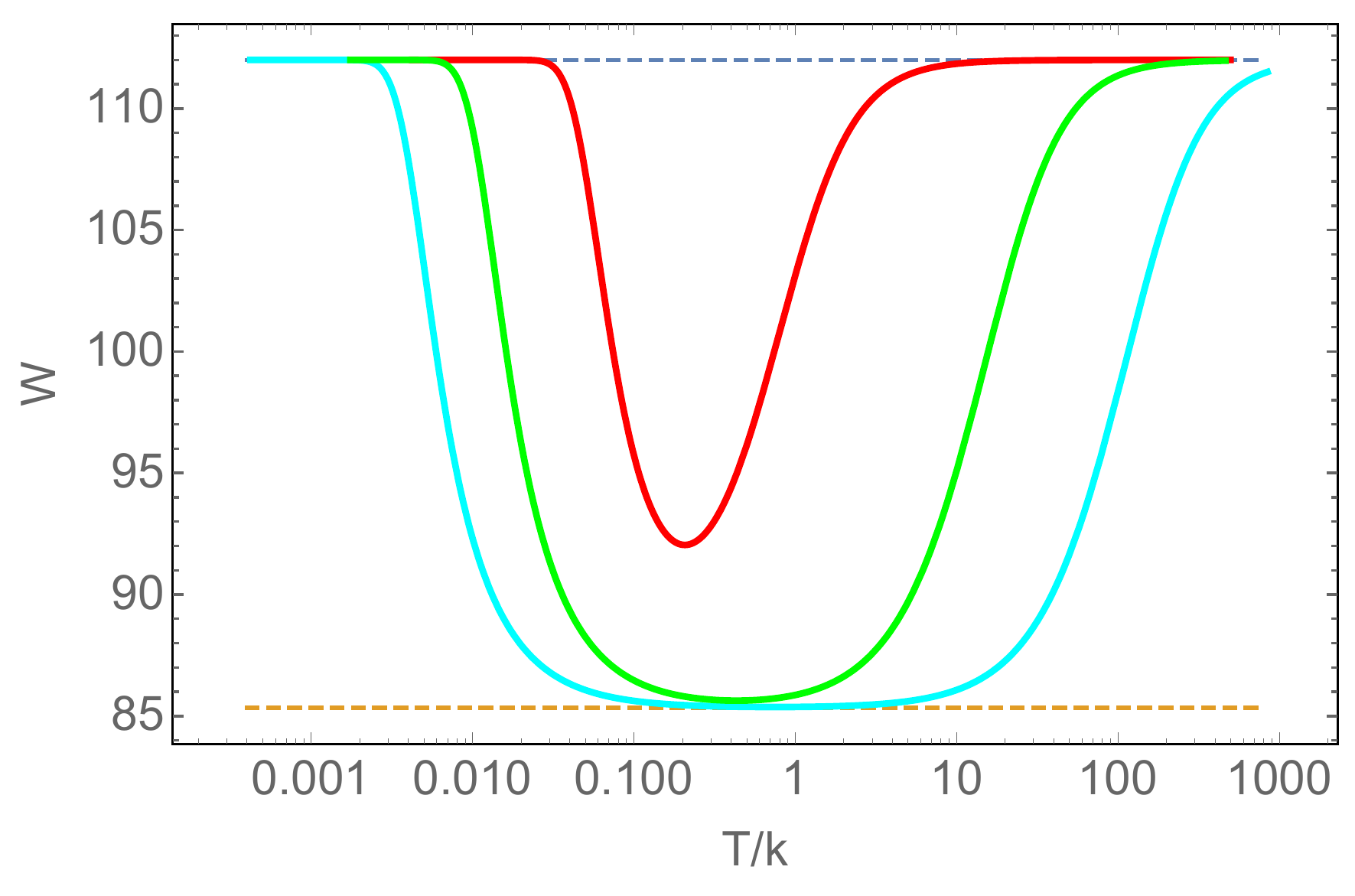}
\caption{Plot of the function $T \,s^{\prime}/s$, where $s$ is the entropy density, and the
Kretschmann scalar at the black hole horizon, $W$, for three values of the deformation parameter $\lambda = 2$ (red), $\lambda = 5$ (green), and $\lambda = 7$ (cyan) (the values of $\lambda$ are increasing from top to bottom starting from the right of the plots). At low temperatures the solutions are approaching domain walls interpolating between $AdS_5$ in the UV and the same $AdS_5$ in the IR. The dashed constant lines in the left panel are at $3$ and at $8/3$, while
in the right panel they are at 112 and 256/3, and the intermediate scaling behaviour, parametrically large in $\lambda$, is clearly revealed for large $\lambda$.
}
\label{fig:scaling}
\end{figure}

We have constructed various black hole solutions for different values of the deformation parameter $\lambda$ using a numerical implementation of the above technique. In figure \ref{fig:scaling} we plot the function $Ts^{\prime}/s$ as a function of $T/k$ for three different values of $\lambda$. Notice that when $Ts^{\prime}/s$ is equal to a constant $\gamma$ the entropy is scaling with temperature according to $s\propto T^{\gamma}$. We also plot the value of $W$, the Kretschmann scalar at the black hole horizon:
\begin{align}\label{kret}
W\equiv [R_{\mu_1\mu_2\mu_3\mu_4}R^{\mu_1\mu_2\mu_3\mu_4}]_{r=r_+}\,.
\end{align}
For each of the three branches we see that when $T/k>>1$,  the entropy scales as $s\propto T^{3}$, as expected.
Indeed, for $T/k>>1$ the temperature scale is much higher than the deformation scale set by $\lambda$ and the solutions are approaching the standard AdS-Schwarzschild solution. This is also confirmed by the value of the Kretschmann scalar at the horizon which is approaching 112, the value for the AdS-Schwarzschild solution.

For $T/k<<1$, we see from figure \ref{fig:scaling} that the solutions behave similarly to the high temperature solutions. We conclude that
at that low temperatures the solutions are approaching domain wall solutions that interpolate between the deformed $AdS_5$ in the UV and the $AdS_5$ vacuum in the IR. For small values of $\lambda$ this was anticipated from our perturbative analysis and now we see that it also occurs for large $\lambda$.

As with other domain walls interpolating between the same $AdS$ space (e.g. \cite{Horowitz:2009ij,Basu:2009vv,Chesler:2013qla,Donos:2013woa,Donos:2014gya}), there can be a renormalisation of relative length scales between the UV and the IR. 
To extract this information we assume that the far IR of the domain wall solution at $T=0$ has a metric as in \eqref{eq:metric_ansatz} with $U=r^2$ and $e^{2V_i}=\bar L^2_ir^2$. Since a rescaling of the radial coordinate in the IR would lead to a rescaling of both the time
coordinate and the spatial coordinates, the invariant quantities, $\bar L_i$, are the relative length scales with respect to the scale fixed by the time coordinate. Since in the UV we approach a unit radius $AdS_5$, the $\bar L_i$ give the renormalisation of the
relative length scales for the RG flow.
 It is slightly delicate to extract the $\bar L_i$ from the finite temperature solutions, because the constants $\mathcal{H}_2$ and $\mathcal{H}_3$
in \eqref{eq:nh_exp} go to zero as $T\to 0$. After considering heating up the putative domain wall
solution with a small temperature and examining the behaviour at the horizon, we deduce that 
the $\bar L_i$ can be obtained by taking the limit
\begin{align}
\bar L_i=\lim_{T/k\to 0} L_i\,,
\end{align}
where we have defined\footnote{In $D$ spacetime dimensions we would have $L_i\equiv\frac{(D-1)}{4\pi T}e^{V_i}|_{r=r_+}$.}
\begin{align}\label{lrenorm}
L_i\equiv
\frac{1}{\pi T}e^{V_i}|_{r=r_+}\,.
\end{align}
Since $e^{V_1}$ and $e^{V_2}$ are the norms of the Killing vectors
$\partial_z$ and $\partial_x,\partial_y$, respectively, we see that we can also write the $L_i$ in a manifestly invariant way as:
\begin{align}\label{lrenorm2}
L_1=2\left(\frac{|\partial_z|}{\kappa}\right )_{r=r_+}\,,\qquad
L_2= 2\left(\frac{|\partial_x|}{\kappa}\right )_{r=r_+}\,,
\end{align}
where here (only) $\kappa$ is the surface gravity of the black hole.

\begin{figure}[t]
\centering
\includegraphics[height=4.7cm]{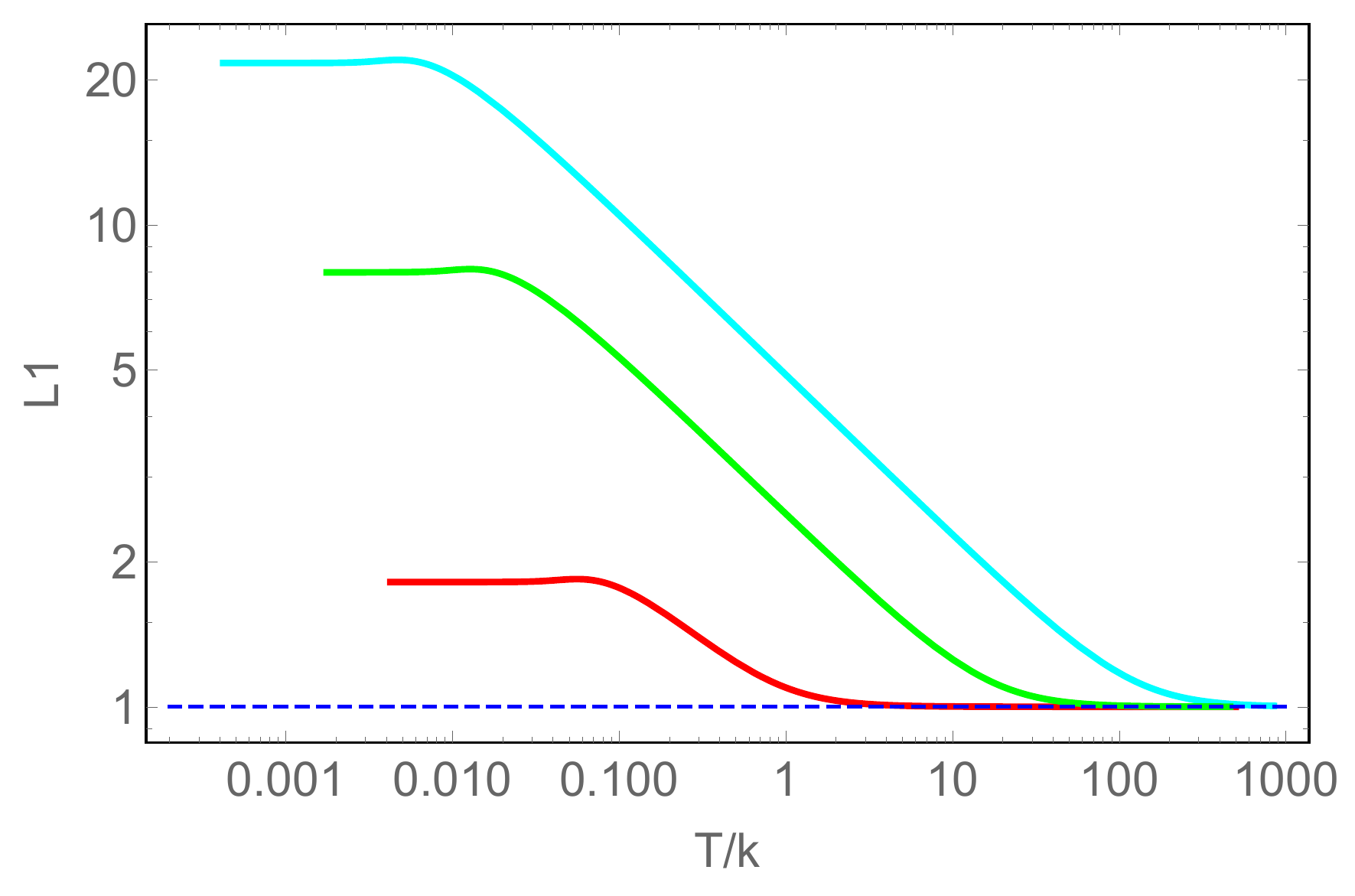}\quad
\includegraphics[height=4.7cm]{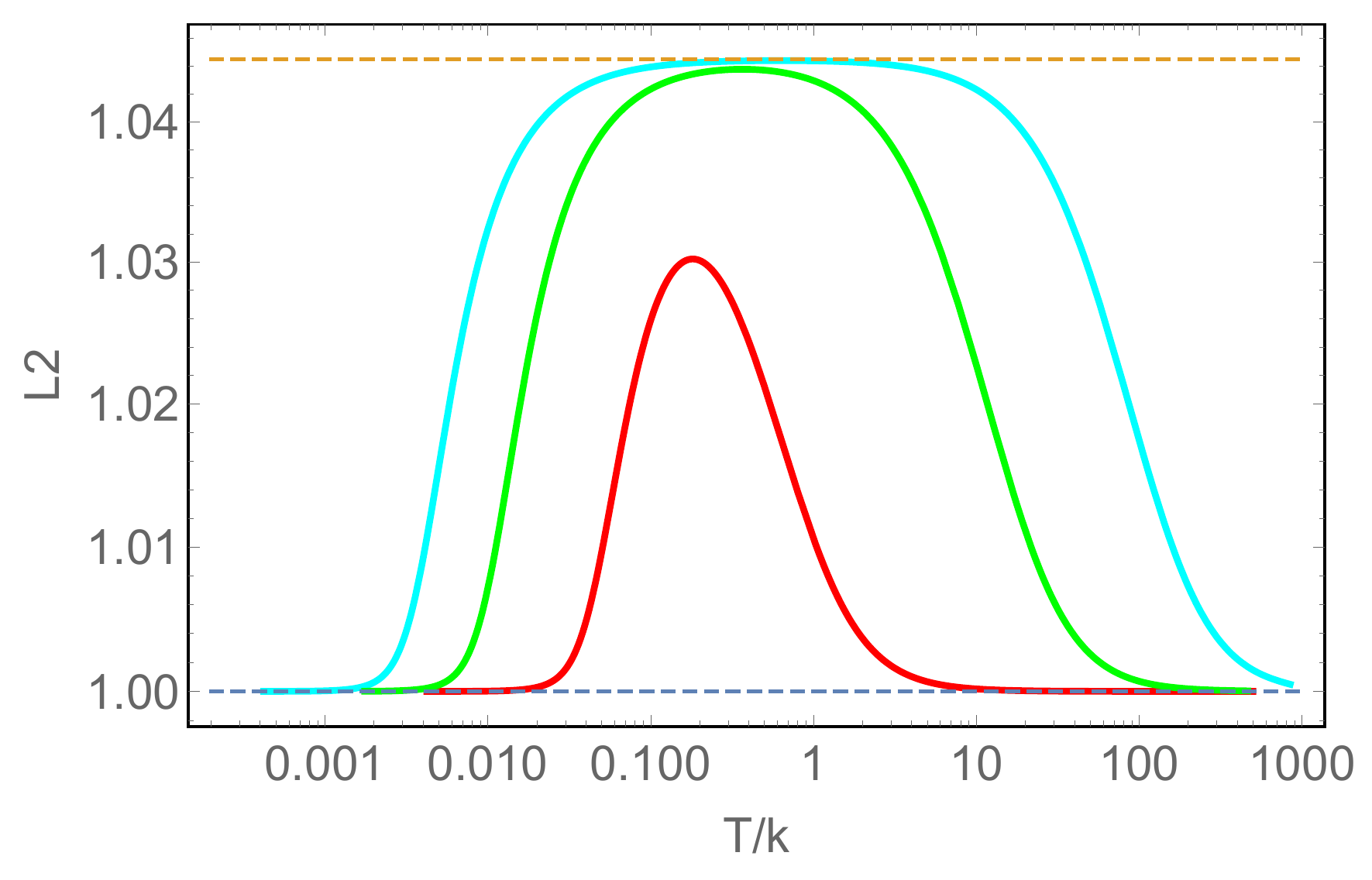}
\caption{Plot of $L_i$, defined in \eqref{lrenorm} against $T/k$ for three values of the deformation parameter $\lambda = 2$ (red), $\lambda = 5$ (green), and $\lambda = 7$ (cyan) (the values of $\lambda$ are increasing from the bottom to the top of the plots). The values of $\bar L_1$ and $\bar L_2$, which are the values of $L_1$ and $L_2$ as $T/k\to 0$, give the relative length renormalisations in the $z$ direction and
the $x,y$ directions, respectively, for the zero temperature $AdS_5$ to $AdS_5$ domain wall solutions.
The dashed orange line at the top of the right plot is at $(12/11)^{1/2}$ associated with the intermediate 
scaling behaviour for large enough $\lambda$.
}
\label{fig:lscale}
\end{figure}

We have plotted $L_i$ as a function of $T/k$ in figure \ref{fig:lscale}. 
We see there is a significant relative length renormalisation in the anisotropic $z$ direction, given by $\bar L_1$, that appears to monotonically increase with $\lambda$. By contrast we see that $\bar L_2=1$. The meaning of this is simply that the domain wall solution at $T=0$ will
preserve the symmetry of $\mathbb{R}^{1,2}$ along the full flow. Indeed it is easy to show that setting $e^{2V_2}=U$ is a consistent
truncation of the equations of motion.

Returning to figure \ref{fig:scaling}, we can also see,
for sufficiently large $\lambda$, the clear emergence of an intermediate scaling behaviour with $s\propto T^{8/3}$, associated with the finite temperature Lifshitz-like scaling solution \eqref{eq:taka_ground} (see \cite{Azeyanagi:2009pr}). 
To quantify this, we note that for the case of $\lambda=7$, for example, our numerics show that the minimum
of the cyan curve in the left plot of \ref{fig:scaling} takes the value $2.6668$ at $T/k=3.733$.
We also see that the scaling region is becoming parametrically
large as $\lambda$ is increased. The lower bound is roughly given by $T/k_{IR}\gtrsim 1$ where $k_{IR}\equiv k/\bar L_1$.
The upper bound satisfies $T/k << e^{\lambda}$ and appears to scale with $e^\lambda$ as anticipated. 
In the intermediate scaling region the value of the Kretschmann scalar at the horizon, $W$, defined in \eqref{kret}, is approaching 256/3 which is the same as that of the black holes associated with heating up the Lifshitz-like scaling fixed point which were presented in equation (2.27) of
\cite{Azeyanagi:2009pr}. It is also interesting to note that in the intermediate scaling regime we
see from figure \ref{fig:lscale} that $L_2\sim (12/11)^{1/2}$. This is precisely the value associated with a domain
wall solution approaching \eqref{eq:taka_ground} at low temperatures.

Finally, for all $T/k$ and for all deformation parameters $\lambda$, we find that
the scalar field $\varphi$ monotonically decreases as a function of the radius from $\lambda$ at $r=\infty$ down to a constant at the black hole horizon. This behaviour can be established by multiplying \eqref{sceom} by $\varphi'$ and then integrating in the radial direction from the horizon to $r$. The integral of the right hand side is positive and after integrating the the left hand side by parts one can establish 
$(\varphi^2)'\ge 0$. Notice, in particular, that the dilaton $e^{\phi}$ is bounded and so string perturbation
theory does not break down in the bulk.

\subsection{Shear viscosity and DC thermal conductivity}
The viscosity shear tensor, $\eta_{ij,kl}$, is defined in terms of the DC limit of the imaginary part of the retarded, two point function of the stress tensor:
\begin{align}
\eta_{ij,kl}=\lim_{\omega\to 0}\frac{1}{\omega}{\mathrm{Im} }G^R_{ij,kl}(\omega)\,,
\end{align}
where $G^R_{ij,kl}(\omega)=\langle T_{ij}(\omega,k=0)T_{kl}(\omega=0,k=0)\rangle$.
The procedure for calculating $\eta_{ij,kl}$ by studying the behaviour of metric perturbations is well known. Here we will import the results for anisotropic holographic lattices presented in \cite{Jain:2015txa} in order to calculate some of the components for our new anisotropic $\tau$-lattice.

We define $\eta_{||}\equiv \eta_{xy,xy}$, which is associated with spin 2 perturbations with respect
to the residual $SO(2)$ rotation invariance in the $x,y$ plane. By examining the behaviour of
the metric perturbation involving $\delta g_{xy}$, as in \cite{Jain:2015txa},
we obtain the standard
result
\begin{align}
\eta_{||}=\frac{s}{4\pi}\,.
\end{align}
We next define $\eta_\perp\equiv \eta_{xz,xz}=\eta_{yz,yz}$, which is associated with spin 1 perturbations with respect to $SO(2)$. 
The equality arises because of the residual $SO(2)$ rotational symmetry in the $x,y$ plane. These components of the
shear viscosity can be obtained by examining metric perturbations involving $h_{x,z}$, $h_{y,z}$ which together carry spin 1 with respect to the $SO(2)$ symmetry. Following the calculation exactly as in \cite{Jain:2015txa}, we obtain
\begin{align}
\eta_\perp&=\frac{s}{4\pi}\,\left. e^{2(V_{2}-V_{1})}\right|_{r=r_{+}}\,.
\end{align}
It is convenient to define the dimensionless quantity
\begin{align}
F=\eta_\perp\,\frac{4\pi}{s}=\left. e^{2(V_{2}-V_{1})}\right|_{r=r_{+}}=\left(\frac{L_2}{L_1}\right)^2\,,
\end{align}
where the $L_i$ were defined in \eqref{lrenorm}.

\begin{figure}[t]
\centering
\includegraphics[height=4.8cm]{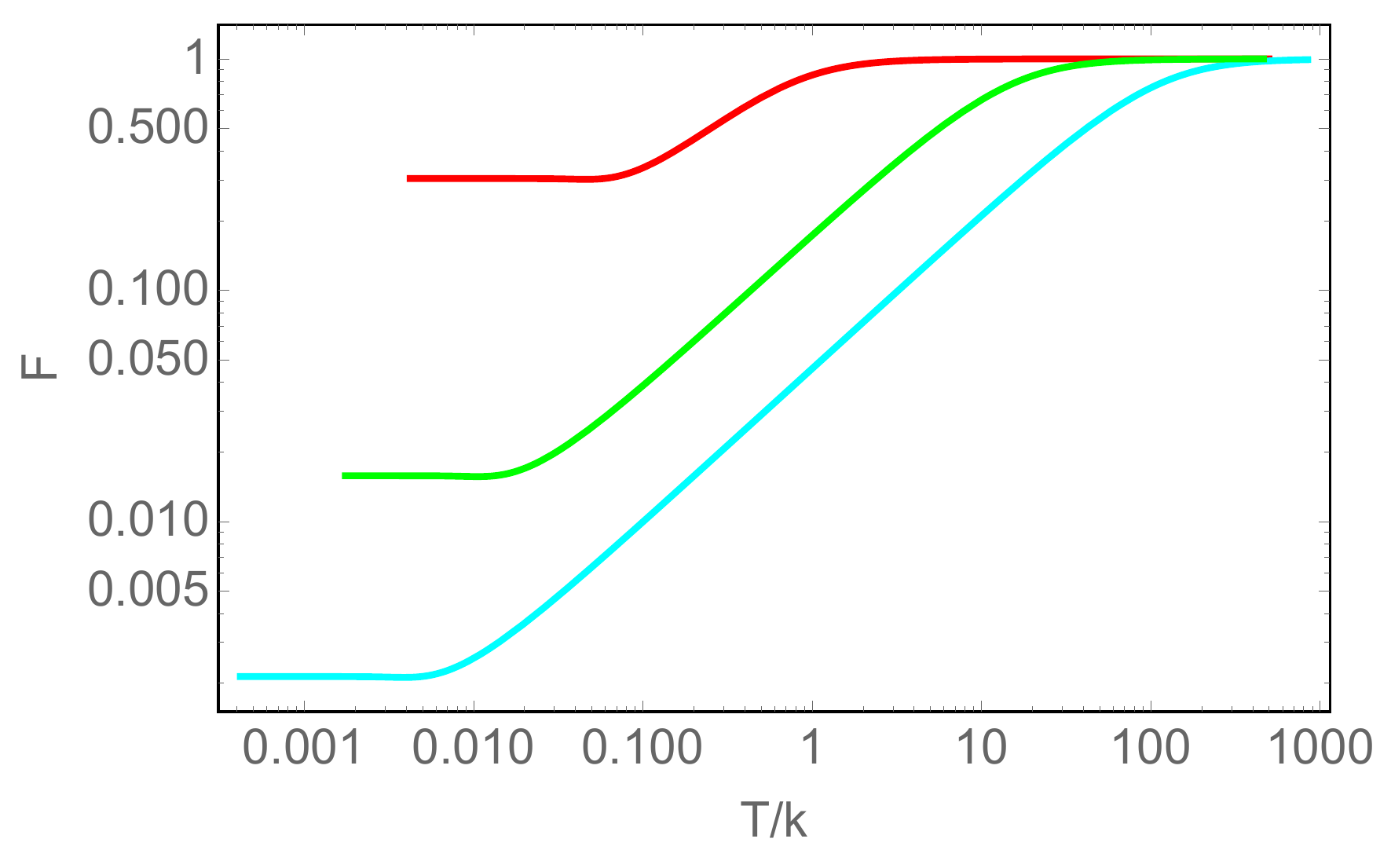}\quad
\includegraphics[height=4.8cm]{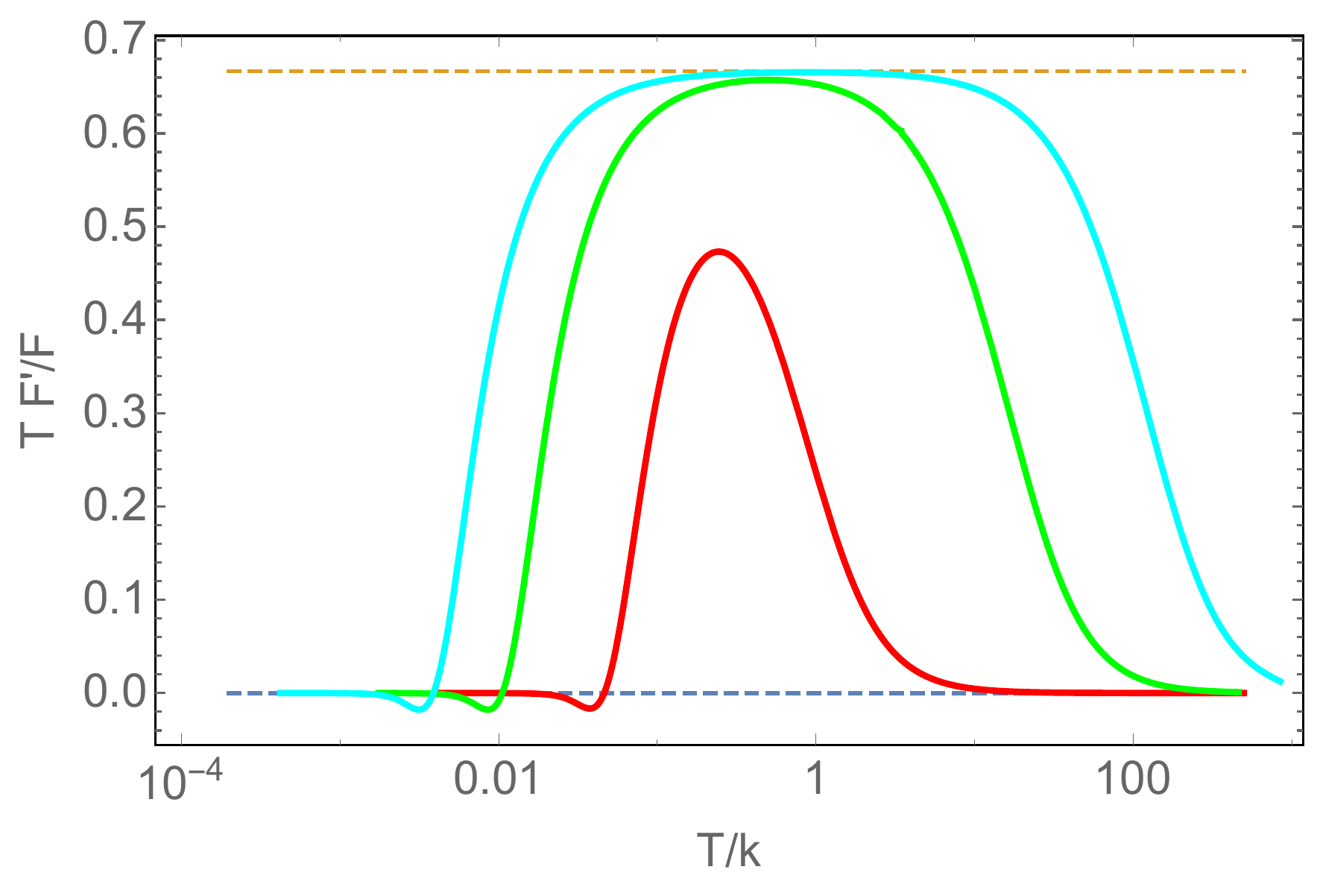}
\caption{Behaviour of $F\equiv \eta_\perp\,\frac{4\pi}{s}$, where $\eta_\perp$ is a component of the 
shear viscosity tensor, for three values of the deformation parameter $\lambda = 2$ (red), $\lambda = 5$ (green), and $\lambda = 7$ (cyan) (the values of $\lambda$ are increasing from top to bottom in the left plot and from
bottom to top in the right plot).
The left plot shows $F$ monotonically decreasing from 1 at high temperatures to a non-zero constant at low temperatures. The right plot shows the intermediate scaling behaviour for large enough $\lambda$,
with the dashed orange line at 2/3.}
\label{fig:scalingF}
\end{figure}

Our numerical results for $F$ are presented in figure \ref{fig:scalingF}.
For $T/k>>1$ our anisotropic solutions approach the standard AdS-Schwarzschild solution and hence in this limit we expect $F\to 1$, as we see in the figure. 

Next, for $T/k<<1$ the solution is approaching a domain
wall solution interpolating between $AdS_5$ in the UV and the same $AdS_5$ in the IR, but with a renormalisation of relative length scales. In this limit we thus expect $F$ to approach a constant, but {\it a priori} it is not clear whether this constant is bigger or smaller than one. Figure \ref{fig:scalingF} shows 
that as $T/k\to 0$ we have\footnote{This behaviour should be contrasted with the low temperature behaviour for
other anisotropic models, where $F$ is vanishing as result of different ground states at $T=0$. 
For example, in the linear dilaton models $F\propto T^2$ \cite{Jain:2014vka}, while for the linear axions solutions of
\cite{Azeyanagi:2009pr,Mateos:2011ix,Mateos:2011tv} we have $F\propto T^{2/3}$ \cite{Rebhan:2011vd,Mamo:2012sy}.}
$F\to F_0$ with $0<F_0<1$. We also see that the value of $F_0$ monotonically decreases with increasing deformation parameter $\lambda$.
Furthermore, as we decrease $T/k$, we see that $F$ monotonically decreases from $F=1$ down to $F=F_0$.

Finally, for large enough $\lambda$, in the intermediate scaling regime we expect that 
$F$ will exhibit the same temperature scaling as for the Lifshitz scaling solution
\eqref{eq:taka_ground}. An inspection of \eqref{eq:taka_ground} reveals that the ratio of length scales imply $F\propto T^{2/3}$. This behaviour is also clearly visible in figure 
\ref{fig:scalingF}.

\begin{figure}[t]
\centering
\includegraphics[height=4.8cm]{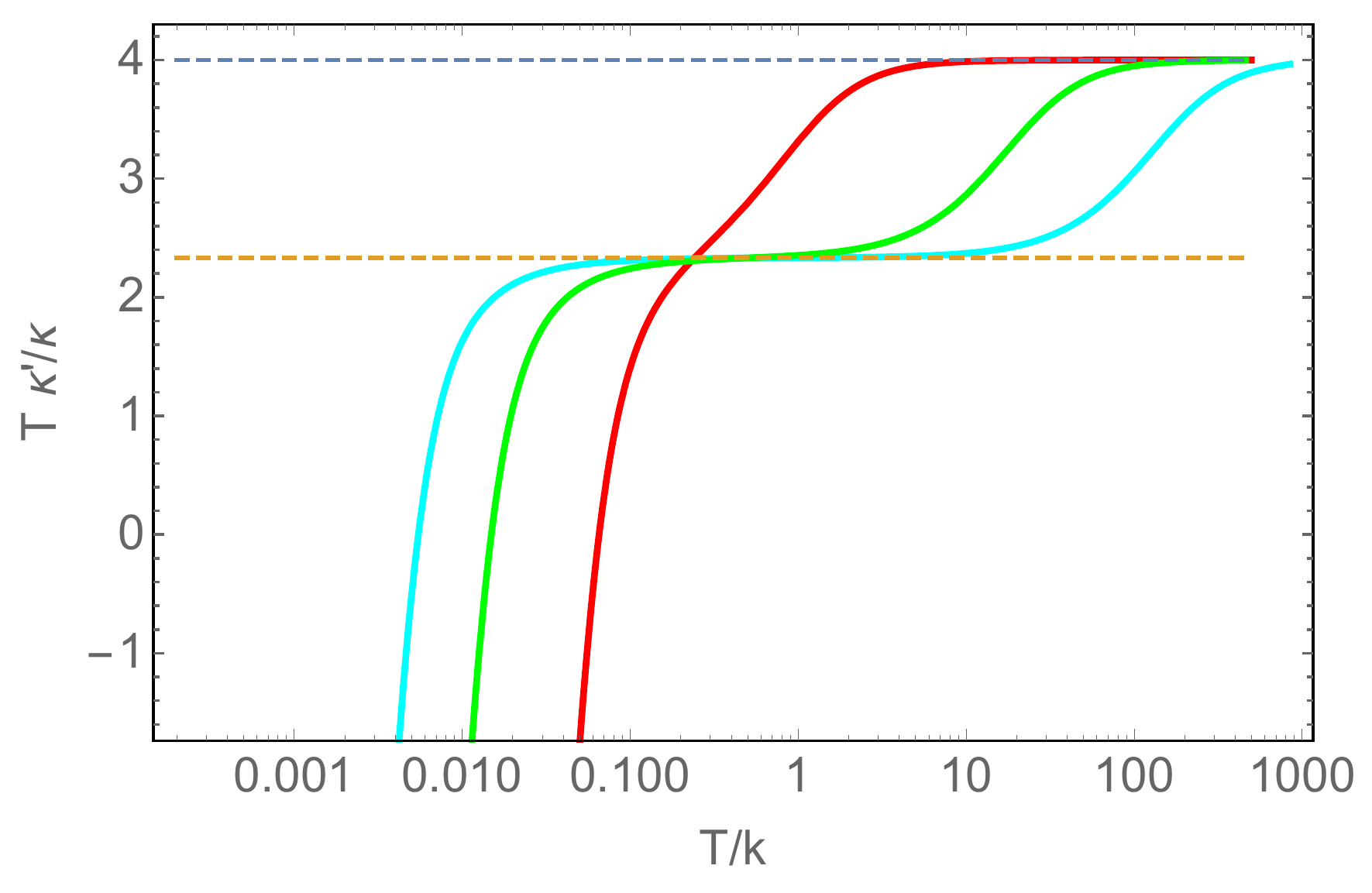}
\caption{Behaviour of the thermal conductivity $\kappa$ as a function of $T/k$
 for three values of the deformation parameter $\lambda = 2$ (red), $\lambda = 5$ (green), and $\lambda = 7$ (cyan)
 (the value of $\lambda$ is increasing from top to bottom starting from the right of the plot).
The dashed blue line is at 4 and the dashed orange line, associated with the intermediate scaling, is at $7/3$.}
\label{kappaplot}
\end{figure}

We conclude this section by calculating the DC thermal conductivity.
The DC thermal conductivity is infinite in the $x$ and $y$ directions, due to translation invariance. However, our $\tau$-lattice breaks translations in the $z$ direction and
hence the conductivity, $\kappa$, in this direction will be finite.
It has been shown that $\kappa$ can be obtained, universally, by solving a system of Stokes equations on the black hole horizon \cite{Donos:2015gia,Banks:2015wha,Donos:2015bxe}. In fact these equations can be solved exactly for holographic lattices that depend on just one spatial direction giving results that were obtained earlier in \cite{Donos:2014cya}. For the case at hand we find that
\begin{align}\label{kapexp}
\kappa=\left. \frac{4\pi s T}{k^2\sinh^2\varphi}\right|_{r=r_{+}}=\frac{16\pi^5 T^4 L_1 L_2^2}{k^2\sinh^2\varphi(r_+)}\,,
\end{align}
and our results are plotted in figure \eqref{kappaplot}.

Given the understanding of the solutions that we have now gained, for $T/k>>1$ we expect that the scaling with temperature of $\kappa$ will
be the same as for the AdS-Schwarzschild solution and hence proportional to $T^4$. Indeed, from
\eqref{kapexp}, with $L_1, L_2\to 1$ and $\varphi(r_+)\to \lambda$ we explicitly see that we have
$\kappa\sim k^2(T/k)^4$ for $T/k>>1$.

Similarly, for $T/k<<1$, we expect that the system will approach an ideal conductor
associated with the momentum dissipating pole for the corresponding Green's function approaching the origin in the complex
frequency plane. From the perturbative solution \eqref{eq:scalar_pert_nh} we can deduce that
\begin{align}
\kappa\sim \frac{128 \pi^7 T^7}{\lambda^2 k^5}e^{2k/\pi T}\,,\qquad T/k<<1,\quad\lambda<<1\,.
\end{align}
More generally by considering heating up a domain wall solution we expect $\kappa\sim (T^7/k^2 k_{IR}^3)e^{2k/\pi T}$ for $T/k<<1$.

Finally, in the intermediate scaling regime, based on the result in \cite{Donos:2014cya}, we expect that
$\kappa\sim k^2(T/k)^{7/3}$. This is exactly the behaviour that our numerics produces as we see in figure \ref{kappaplot}.
For $T<<k$ the Boltzmann behaviour of the thermal resistivity, $\kappa^{-1}$, can be understood as a consequence of the absence of low-energy excitations supported at the lattice momentum $k_{IR}$. More precisely, if we denote by $\cal O$ the operator dual to the axion and dilaton then the behaviour is a consequence of the vanishing of the
spectral function $Im G^R_{\cal {O}\cal {O}}(\omega, k_{IR})$ as $\omega\to 0$, as explained in 
\cite{Hartnoll:2012rj}.

All of the above features for $\kappa$ are clearly displayed in figure \ref{kappaplot}.

\section{Discussion}\label{disc}
We have investigated a new class of anisotropic plasmas associated with
the infinite class of CFTs that have $AdS_5\times X_5$ holographic duals. The 
plasmas arise from periodic deformations of the axion and dilaton of type IIB supergravity
that depend on just one
of the spatial directions. While these deformations do not modify the far IR physics, apart from
a simple renormalisation of relative lengths scales,
for sufficiently large deformations there is a novel intermediate
scaling regime governed by a Lifshitz-like solution with a linear axion that was found in \cite{Azeyanagi:2009pr}.

The deformations that we have considered arise from a distribution of $D7$-branes
and anti- $D7$-branes smeared along one of the spatial directions of the field theory.
It is rather remarkable that one can construct back-reacted solutions for such configurations. It is also suggestive that the solutions may suffer from instabilities and it would be worthwhile to investigate this issue in more detail. 

For the particular case of $X_5=S^5$, associated with $N=4$ Yang-Mills theory, it is known that the Lifshitz-like solution is unstable \cite{Azeyanagi:2009pr}. At finite temperature it was shown that the linear axion solutions have phase transitions
\cite{Banks:2015aca,Banks:2016fab} leading to new branches of solutions. It would be interesting to investigate whether similar instabilities and phase transitions
occur for the $\tau$-lattices we have constructed here. It seems plausible that there is a critical value of the deformation parameter where instabilities set in.
For the case of the linear axion deformations, the addition of a gauge-field has been investigated in \cite{Cheng:2014qia, Banks:2016fab}. Similarly incorporating a gauge field with the new $\tau$-lattices is another topic for further study.

The deformations that we have constructed are periodic in the spatial direction. 
Indeed as one moves along a period in the spatial direction the field configuration traverses a circle in the Poincar\'e disc.
One can construct other periodic configurations by utilising the
exact $SL(2,Z)$ symmetry of type IIB string theory. More precisely, as one moves
along a period in the spatial direction, one can demand that while $\tau$ itself is not
periodic, it is periodic after acting with a non-trivial element of $SL(2,Z)$. Examples of such solutions can easily be obtained from the solutions we have presented here by taking $SL(2,Z)$
quotients of the circle on the Poincar\'e disc. Notice that integrating along the periodic spatial direction would then lead to a net $(p,q)$ D7-brane charge. There are many more possibilities when additional spatial directions are involved and it would be interesting to explore such constructions in more detail.

\section*{Acknowledgements}

The work of JPG is supported by STFC grant ST/L00044X/1, 
EPSRC grant EP/K034456/1,
and by the European Research Council under the European Union's Seventh Framework Programme (FP7/2007-2013), ERC Grant agreement ADG 339140. JPG is also supported as a KIAS Scholar and
as a Visiting Fellow at the Perimeter Institute. 
In addition OSR is supported by CONACyT.

\appendix
\section{Q-lattices and $SL(2,R)$ conjugacy classes}\label{appa}

The holographic lattices of \cite{Azeyanagi:2009pr,Mateos:2011ix,Mateos:2011tv,Jain:2014vka}, and the ones studied here, are all examples of Q-lattices \cite{Donos:2013eha} in which one exploits the fact that the scalars parametrise the group manifold
$SL(2,R)$ and consequently the gravity model admits a global $SL(2,R)$ symmetry.
In particular, the spatial dependence of the scalars on the $z$ direction is given by a specific orbit of the $SL(2,R)$ action. There are three distinct cases to consider corresponding
to the three different conjugacy classes of $SL(2,R)$. 

Write a general $SL(2,R)$ matrix as
\begin{equation}
{\cal M}=\begin{pmatrix}
a&b\\
c&d\\
\end{pmatrix},\qquad ad-bc=1\,.
\end{equation}
The three conjugacy classes are determined
by the trace: the parabolic class has $|Tr{\cal M}|=2$, the hyperbolic class
has $|Tr{\cal M}|>2$ and the elliptic class has $|Tr{\cal M}|<2$.

The $SL(2,R)$ action on $\tau=\chi + i e^{-\phi}$ 
is given by $\tau\to (a\tau+b)/(c\tau+d)$. Suppose we start with $\chi=0$ and $\phi=\phi_0$. Then acting with the one-parameter family of $SL(2,R)$ transformations in the parabolic conjugacy class given by
\begin{equation}
{\cal M}=\begin{pmatrix}
1& kz\\
0&1\\
\end{pmatrix},
\end{equation}
induces $\chi\to kz$ with $\phi_0$ unchanged. This generates the ansatz for scalar fields given by $\phi=\phi(r)$
and $\chi= kz$ that was used in \cite{Azeyanagi:2009pr,Mateos:2011ix,Mateos:2011tv}.

Next we consider the transformations in the hyperbolic conjugacy class of the form
\begin{equation}
{\cal M}=\begin{pmatrix}
e^{-kz/2}&0\\
0&e^{k z/2}\\
\end{pmatrix}\,.
\end{equation}
This induces $\phi_0\to \phi_0+kz$ with $\chi=0$ unchanged. 
This generates the ansatz $\phi=kz$ and $\chi=0$ used in \cite{Jain:2014vka}.

Finally, we consider the transformations in the elliptic conjugacy class of the form
\begin{equation}\label{elliptic}
{\cal M}=\begin{pmatrix}
\cos(\frac{kz}{2})&\sin(\frac{kz}{2})\\
-\sin(\frac{kz}{2})&\cos(\frac{kz}{2})\\
\end{pmatrix}\,.
\end{equation}
After writing $\phi_0=\varphi_0$ we find that this induces the transformation
\begin{align}\label{eq:field_redef2}
\chi&\to\frac{\sinh\varphi_0 \, \sin kz}{\cosh\varphi_0+\sinh\varphi_0\,\cos kz}\,,\notag \\
e^\phi&\to\cosh\varphi_0+\sinh\varphi_0\,\cos kz\,.
\end{align}
This generates $\varphi=\varphi(r)$, and  $\alpha=k\,z$, exactly as in
\eqref{eq:field_redef} and \eqref{eq:q_ansatz}. Notice that as we traverse once in the Poincar\'e disc via 
$z\to z+2\pi/k$, we see from \eqref{elliptic} that we move from ${\cal M}=1$ to ${\cal M}=-1$. In other words we have a closed orbit in $PSL(2,R)$.

\providecommand{\href}[2]{#2}\begingroup\raggedright\endgroup

\end{document}